%Paper: hep-th/9305018
%From: JXLU@crnvma.cern.ch
%Date: Thu, 06 May 93 15:07:19 SET

%
%input harvmac
%

\input harvmac.tex

%%%%%%%%%%%%%%%%%%%%%%%%%%REFERENCES FOR THE DISSERTATION%%%%%%%%%%%%%%%%%%%%%%

\lref\berst {E. Bergshoeff, E. Sezgin and P. Townsend, Phys. Lett. {\bf B189}
 (1987)  75.}

\lref\achetw {A. Achucarro, J. Evans, P. Townsend and D. Wiltshire, Phys.
Lett. {\bf B198}  (1987)  441.}

\lref\duflfb {M. J. Duff and J. X. Lu, Nucl. Phys. {\bf B354}  (1991)  141.}

\lref\dabghr {A. Dabholkar, G. W. Gibbons, J. A. Harvey and F. Ruiz Ruiz,
Nucl. Phys. {\bf B340}  (1990)  33.}

\lref\dufhis {M. J. Duff, P. S. Howe, T. Inami and K. Stelle, Phys. Lett.
{\bf B191}  (1987)  70.}

\lref\dufs {M. J. Duff and K. Stelle, Phys. Lett. {\bf B253}  (1991)  113.}

\lref\calhswb {C. Callan, J. Harvey and A. Strominger, Nucl. Phys. {\bf B367}
 (1991) 60.}

\lref\duflthb {M. J. Duff and J. X. Lu, Phys. Lett. {\bf B273}  (1991)  409.}

\lref\hors {G. Horowitz and A. Strominger, Nucl. Phys. {\bf B360}  (1991)
197. }

\lref\duflpb {M. J. Duff and J. X. Lu, ``Black and super $p$-branes in diverse
dimensions", CERN-TH.6675/92 and CTP--TAMU--54/92  (1992).}

\lref\duflsds{M. J. Duff and J. X. Lu,``The self-dual Type IIB superstring"
 (unpublished).}

\lref\duflbscan{M. J. Duff and J. X. Lu, ``Type II $p$-branes: The Brane-Scan
Revisited", CERN-TH.6560/92 and CTP/TAMU-37/92 (to appear in Nucl. Phys.
{\bf B}).}

\lref\stra{J. Strathdee, Int. J. Mod. Phys. {\bf A2} (1987) 273.}

\lref\salsone{A. Salam and E. Sezgin, ``Supergravities in Diverse Dimensions",
North-Holland and World Scientific.}

\lref\luthesis{J. X. Lu, ``Supersymmetric Extended Objects", Texas A\&M,
Ph.D. Thesis (1992).}

\lref\salstwo {A. Salam and E. Sezgin,  Nucl. Phys. {\bf B258} (1985) 284.}

\lref\sen{A. Sen,  Phys. Lett. {\bf B274} (1992) 34.}

\lref\hass{S. F. Hassan and A. Sen,  Nucl. Phys. {\bf B375} (1992) 103.}

\def\dg{\hbox{$^\dagger$}}

\def\at{\hbox{$^{\ast}$}}

\def\half{{1\over2}}

\Title{\vbox{\baselineskip12pt \hbox{CERN-TH.6691/93}}}
{\vbox{\centerline{\bf On the Determination of the Dilaton--Antisymmetric
Tensor Couplings}\vskip2pt \centerline{\bf in  Supergravity Theories}}}
\baselineskip=18pt
%For more complicated situations, substitute for {\it either\/} argument:
%\Title{\vbox{\baselineskip12pt\hbox{HUT$p$-88/A000}\hbox{SLAC-PUB 88-001}
%               \hbox{photocopy at own risk}}}
%{\vbox{\centerline{This title is too long to fit}
%       \vskip2pt\centerline{comfortably on one line*}}}
%   \footnote{}{*optional footnote on title}

\centerline {J. X. Lu{\footnote\dg{Supported partially by
a World Laboratory Fellowship}}}
\centerline{\it{CERN, Theory Division, CH-1211, Geneva 23, Switzerland}}
\centerline{jxlu@cernvm.cern.ch}
\bigskip
\centerline{\bf ABSTRACT}

A new approach is provided to determine the dilaton--antisymmetric tensor
coupling in a supergravity theory by considering the static supersymmetric
field
configuration around a super extended object, which is consistently formulated
in a curved
superspace. By this, the corresponding SUSY transformation rules can also be
determined for vanishing fermionic fields as well as bosonic fields other than
those in the determined coupling. Therefore, we can, in turn, use this
determined
part of the supergravity theory to study all the related vacuum-like solutions.
We have determined  the dilaton--antisymmetric tensor couplings, in which
each of the antisymmetric tensors is a singlet of the automorphism group of the
corresponding superalgebra, for every  supergravity multiplet.
This actually happens only for $N \leq 2$ supergravity theories, which agrees
completely with the spin-content analysis and the  classified
$N \leq 2$ super $p$-branes, therefore giving more support to the existence of
the fundamental Type II $p$-branes. A prediction is made of the $D = 9, N = 2$
supergravity which has not yet been written down so far.

\bigskip
\noindent
CERN-TH.6691/93

\Date{May 1993}
\eject

\newsec{\bf Introduction}

Recently, a lot of attention has been given to finding
Reissner--Nordstr\"om-like and super extended solutions from certain
supergravity theories
\refs{\dabghr, \duflfb, \dufs, \calhswb, \duflthb, \hors, \sen, \hass,
\duflpb}.
In these solutions,
the dilaton--antisymmetric tensor couplings
of the corresponding supergravity theories play essential roles. Without
 this coupling,  the dilaton field is as trivial as a scalar field in
ordinary general relativity for the aforementioned solutions. Therefore,
determining this coupling is essential at least for charged classical
solutions. Supergravity multiplets in diverse dimensions have all been
classified some time ago (e.g. see \stra). Almost all the corresponding
actions or equations of motion have  been given either by direct constructions
or by dimensional reductions
and truncations from $D = 11$ or $D = 10$ Type IIB supergravity (see e.g.
\salsone). Therefore, in
principle, we should know all these couplings in the corresponding supergravity
theories, which are determined by {\it space--time supersymmetries}. In this
note, we provide a simple approach to determine this
coupling from considering the static supersymmetric field configuration
produced by a
super extended object consistently coupled with background fields, which is
exactly opposite
to the current direction, i.e. finding  super extended solutions from
a known supergravity theory. This study is motivated by
 recent work \duflpb. The physical motivation here is: if a source is
space--time
supersymmetric, what else  could the fields around the source  be but
supersymmetric. From this, we
must have a correspondence between a super extended object and a
certain supergravity multiplet (possibly reducible), which was also discussed
in \achetw\ for
some other reasons. Although, unlike in string cases, we are not sure that a
super $p$-brane (for $p > 1$)  has  its background fields, i.e. the
corresponding
supergravity multiplet, as its zero modes, classically everything is perfect
for formulating the  super $p$-brane in a curved superspace, since this only
requires that the superspace $(p + 2)$-form $H$ and the superspace torsion $T$
satisfy certain  constraints \refs{\achetw, \berst}. These constraints,
 in general, contain but may not be sufficient to give the field equations of
the corresponding supergravity theories, except for the $D = 11$
 and $D = 10, N = 1$ cases \refs{\achetw, \berst}.  It is also worth pointing
out that as for Green--Schwarz-like super $p$-brane, the space--time
supersymmetry
and the world-volume supersymmetry are linked by the so-called
$\kappa$-symmetry, a local fermionic gauge symmetry. Given a super extended
object consistently formulated in a curved superspace,
 the field configuration produced by the object  is  described
 by a supersymmetric theory, therefore the dilaton--antisymmetric tensor
coupling must be determined from this field configuration.
 Once  this coupling is determined, we
can use the corresponding action to find the Reissner--Nordstr\"om-like
solution. We then find the supersymmetric solution by taking the mass =
charge limit since we know that we begin with a supersymmetric theory.
Using this supersymmetric solution, we can determine the supersymmetric
transformation rules for vanishing gravitino and dilatino. By this
approach, we can determine a supergravity theory up to the stage that
is good for those vacuum-like solutions related to the determined couplings.
It  also provides a
primary step for determining a complete supergravity theory. Up to present
time, only Type I super $p$-brane actions for $p > 1$ \refs{\achetw,\berst}
and   superstring actions of both Type I
and II are given, although there might exist  Type II
super $p$-branes for $p > 1$ \duflbscan. All these known super $p$-brane
actions
have manifest both space--time and world-volume supersymmetries, which is
crucial to the approach
discussed in this paper. In what follows, we  use just these supersymmetric
actions as examples, to demonstrate that the above is correct.
Moreover, our results apply also to those fundamental Type II $p$-branes and
to some other supersymmetric extended solitons  classified in \duflbscan.
Various implications of this are discussed in detail.

\newsec{\bf Scaling arguments}

{}From our experience, we know that the fields involved in a
Reissner--Nordstr\"om-like or the charge = mass limit extended solution from
a supergravity theory are the graviton, the dilaton and an
antisymmetric tensor. The most general action for these fields written in
canonical variables with standard normalization factors are
\eqn\genact{I_D = {1\over 2 \kappa^2}\int d^D x \sqrt {- g} \bigg[R - \half
(\partial \phi)^2 - {1\over 2 (d + 1)!} f(\phi) F_{d + 1}^2 \bigg],}
where $D$ is the space--time dimension, $\kappa$ the $D$-dimensional
gravitational constant, $f(\phi)$ the undetermined coupling, and $F_{d + 1}$
the field strength of a  $d$-form potential $B_d$, which is given by
\eqn\poten{F_{d + 1} = d B_d .}

We now consider \genact\ as describing the field configuration produced by
a bosonic $(d - 1)$-brane source  sitting in the same space--time
whose world-volume action is
\eqn\pbact{\eqalign{S_d = - T_d \int d^d \xi \bigg[&\half \sqrt {-\gamma}
\gamma^{ij} \partial_i X^M \partial_j X^N G_{MN} -
{d - 2\over 2}\sqrt {-\gamma}\cr
&+{1\over d!}\epsilon^{i_1 \cdots i_d}\partial_{i_1} X^{M_1} \cdots
\partial_{i_d} X^{M_d} B_{{M_1}\cdots {M_d}}\bigg],\cr}}
where $M = 0, \cdots, D - 1$, $T_d$ the $(d - 1)$-brane tension, $G_{MN}$ the
$(d - 1)$-brane $\sigma$-model metric and $\gamma_{ij}$ the induced
world-volume
 metric, which is given by
\eqn\induced{\gamma_{ij} = \partial_i X^M \partial_j X^N G_{MN}.}
Actually, \pbact\ is the bosonic sector of the corresponding super $p$-brane
action, and we  postpone the discussion of supersymmetry to the next section.
For present purposes, the action \pbact\ is enough. We consider the following
general scalings of fields $G_{MN}$ and $B_{M_1 \cdots M_d}$ in \pbact,
\eqn\scaling{\eqalign{G_{MN} &\rightarrow \lambda^2 G_{MN}\cr
          B_{M_1\cdots M_d} &\rightarrow \beta B_{M_1 \cdots M_d}.\cr}}
Under these scalings, $S_d \rightarrow \lambda^d S_d$ provided we take
$\beta = \lambda^d$. Hence,  $S_d$ scales homogeneously. If \genact\ does
describe the field configuration, we expect that
$I_D \rightarrow \lambda^d I_D$, i.e. scales the same way as $S_d$ under the
same scalings. From this requirement and the metric relation
$G_{MN} = \Omega(\phi)g_{MN}$ as discussed in \luthesis, we have
\eqn\fomega{\eqalign{f(\phi + c) &= \lambda^{- {2 d \tilde d \over D - 2}}
f(\phi),\cr
\Omega (\phi + c) &= \lambda^{{2\tilde d \over D - 2}} \Omega (\phi ),\cr}}
where the constant shift $c$ of the dilaton field corresponds to the rescaling
$\lambda$ of the metric and $c = 0$ to $\lambda = 1$,
and where $\tilde d = D - 2 - d$. If there is no dilaton field, i.e. setting
$\phi = 0$ in the above theory, we should have $f(\phi = 0) = 1$, which gives
the
standard field strength kinetic term, and $\Omega (\phi = 0) = 1$ since now
the canonical metric $g_{MN}$ is identical to the $\sigma$-model metric
$G_{MN}$. Noticing these facts, it is not difficult to solve the above
equations
as $\lambda = 1 + \epsilon$, where $\epsilon$, and therefore $c$, is
infinitesimal. The results are
\eqn\sfomega{\eqalign{f(\phi) &=  e^{- \alpha (d) \phi},\cr
                      \Omega (\phi ) &= e^{{\alpha (d) \over d}\phi}, \cr}}
where $\alpha (d)$ is given by
\eqn\expalpha{\alpha (d) = {2 d \tilde d\over D - 2} {\epsilon\over c}.}
For future reference, we explicitly list  the scalings of the canonical
metric $g_{MN}$, the dilaton and the antisymmetric potential
$B_{M_1 M_2 \cdots M_d}$ as
\eqn\genscalings{\eqalign{g_{MN} &\rightarrow \lambda^{2 d\over D - 2} g_{MN},
\cr
B_{M_1 M_2 \cdots M_d} &\rightarrow \lambda^d B_{M_1 M_2 \cdots M_d}, \cr
e^\phi &\rightarrow \lambda^{2 d \tilde d \over \alpha (d) (D - 2)}
e^\phi.\cr}}
By simple scaling arguments, we have determined the relation between
canonical metric and $\sigma$-model metric,  and the action \genact\ as
follows
\eqn\metrelation{G_{MN} = e^{{\alpha (d)\over d}\phi} g_{MN},}
\eqn\genactone{I_D = {1\over 2 \kappa^2}\int d^D x \sqrt {- g} \bigg[R - \half
(\partial \phi)^2 - {1\over 2 (d + 1)!} e^{{- \alpha (d)}\phi}
F_{d + 1}^2 \bigg].}

\newsec{\bf World-volume approach to super $p$-brane field configurations}

In this section, we will use the world-volume approach to discuss a static
super $p$-brane field configuration.  Given a super $p$-brane world-volume
action
formulated consistently in a curved superspace, a static super $p$-brane
configuration, which is the most special one, must satisfy the world-volume
equations of motion, the supersymmetry transformations, and the superspace
constraints on the background fields. The constraints are actually and
should be (for the reasons given in section 1) those of certain supergravity
theories as formulated in superspace language. In general, they are not
sufficient
to give all the equations of motion of the corresponding supergravity,
except for
a very few cases. Even for these cases, extracting the equations of motion
from the corresponding superspace constraints is a very complicated process.
 In what follows,
we will demonstrate that  a static super $p$-brane
field configuration does satisfy the world-volume equations of motion and
supersymmetry transformation rules. We are not going to attack the superspace
constraints; instead, by a simpler approach, we require that the proposed part
\genactone\ of the bosonic sector of the corresponding supergravity action
describe the same super $p$-brane field configuration. By doing this,
we determine completely
 not only the unknown parameter $\alpha (d)$ in \expalpha, thus the action
\genactone, but also the static field configuration.
 We will consider, as examples,
 the only known covariant super $p$-brane actions with manifest ($N = 1$ for
$p > 1$ and $N = 1, 2$ for $p = 1$) space--time supersymmetries. Certainly,
our approach is not limited to these actions, its applicability will be given
in section 5.
 The Green--Schwarz-like
super $(d - 1)$-brane action \refs{\achetw, \berst}  is
\eqn\spbact{\eqalign{S_d &= - T_d\int d^d \xi \bigg (\half \sqrt {- \gamma}
\gamma^{ij} E_i\,^A E_j\,^B \eta_{ A B} -
{d - 2\over 2} \sqrt{- \gamma}\cr
&\quad +{1\over d} \epsilon^{i_1 i_2 \cdots i_d} E_{i_1}\,^{{\hat A}_1}
E_{i_2}\,^{{\hat A}_2}\cdots E_{i_d}\,^{{\hat A}_d} B_{{\hat A}_d \cdots
{\hat A}_1}\bigg),\cr}}
where $\xi^i$ ($i = 0, 1, \cdots, d - 1$) are coordinates for the
$d$-dimensional world-volume with metric $\gamma_{ij}$, swept out by the
(closed) $(d - 1)$-brane in the course of its evolution. The
$E_i\,^{\hat A} ({\hat A} = A, \alpha)$ are defined
by
\eqn\defE{E_i\,^{\hat A} = \partial_i z^{\hat M} E_{\hat M}\,^{\hat A},}
where $\partial_i \equiv \partial/\partial \xi^i$,
$z^{\hat M} = (X^M, \theta )$ are the coordinates of $D$-dimensional superspace
($M = 0, 1, \cdots, D - 1$) and $E_{\hat M}\,^{\hat A}$ is the supervielbein.
Our metric convention for both $d$ and $D$ dimensions is ``mostly plus", i.e.
($-, +, +, \cdots, +$). Since action \spbact\ is a scalar of the super target
space, it is invariant under space--time supersymmetry transformation. Besides,
it has a very important local fermionic gauge symmetry, the so-called
$\kappa$-symmetry, which is, in terms of
$\delta z^{\hat A} \equiv \delta z^{\hat M} E_{\hat M}\,^{\hat A}$,
\eqn\skappa{\eqalign{&\delta z^A = 0, \qquad \delta z^\alpha =
( 1 + \Gamma)^\alpha\,_\beta \kappa^\beta (\xi),\cr
&\Gamma^\alpha\,_\beta \equiv {(- 1)^{(d - 2)(d + 1)/4}\over d!}
{\epsilon^{i_1 i_2 \cdots i_d}\over \sqrt {- \gamma}} E_{i_1}\,^{A_1} \cdots
E_{i_d}\,^{A_d} (\Gamma_{A_1 \cdots A_d})^\alpha\,_\beta , \cr}}
where the parameter  $\kappa^\beta$ is a world-volume scalar but space--time
spinor, and $\Gamma_{A_1 \cdots A_k}$ is the antisymmetrized product of
$k$ (space--time) $\Gamma$-matrices $\Gamma_A$. This symmetry, along with the
space--time supersymmetry, guarantees the world-volume supersymmetry.
Demanding
this symmetry gives some constraints on the super background fields, which
are actually the superspace constraints of the corresponding supergravity
multiplet \berst. For $D = 11$ or $D = 10, N =1$, these
constraints are equivalent to the equations of motion of the corresponding
supergravity  \refs{\achetw, \berst, \dufhis}. The superspace $d$-form
potential
$B_d$ is defined as
\eqn\bdef{B_d = {1\over d!} E^{{\hat A}_1} \cdots E^{{\hat A}_d} B_{{\hat A}_d
\cdots {\hat A}_1},}
and its field strength is given by $F_{d + 1} = d B_d$. The constraint on
$F_{d + 1}$ requires that it satisfy the superspace Bianchi identity
\eqn\sbianchi{d F_{d + 1} = 0.}
A classification of these super $p$-branes with the allowed $(p, D)$ values was
given in \achetw\ for $D - 1> p \geq 1$. Besides, the number $N$ of space--time
supersymmetries is restricted to 1 for $p> 1$, while $N = 1, 2$ are both
possible for $p = 1$. These super $p$-branes consist of the four sequences
labelled
by ${\bf R, C, H, O}$ (the four composition-division algebras, for reasons
explained in \achetw).
By inspection one can verify that for each super $p$-brane there is a
supergravity
multiplet (possibly reducible) containing a $(p + 1)$-form gauge field $B$ with
a $(p + 2)$-form field strength $F = d B$, i.e. the one in \genactone\ for
$d = p + 1$. However, in what follows, we are interested in only {\bf O} and
{\bf H} sequences. The $(p + 2)$-form gauge field in sequences {\bf R} and
{\bf C} is either the dilaton itself, or some auxiliary
field, or some scalar field in the matter supermultiplet as discussed in
\achetw. This does not give the coupling that is important for the purpose of
 this
paper, although it may be useful for some other intentions. Actually, it gives
$\tilde d = D - 3 - p \leq 0$ and is out of the  condition $\tilde d \geq 1$
in the following analysis.
For $p> 1$ the equation of motion for $\gamma_{ij}$ is
\eqn\wmetric{\gamma_{ij} = E_i\,^A E_j\,^B \eta_{AB}.}
(For $p = 1, \gamma_{ij}$ is determined only up to an arbitrary scale factor.)
Using this equation one can show that the $\Gamma$ in \skappa\ satisfies
$\Gamma^2 = 1$, hence $\half (1 + \Gamma)$ is a projection operator. For this
reason, only {\it half} the components of $\kappa^\beta$ are effective in the
gauge transformation \skappa\ which can be used to gauge away {\it half}
of the components of $\theta$. After this short review, we are now ready to
discuss a static super $(d - 1)$-brane field configuration. As usual, we set
the fermionic coordinate $\theta = 0$, then the super target space is just
usual space--time and the super ($d - 1$)-action \spbact\ reduces to its
bosonic sector
\pbact. In order to have a super $(d - 1)$-brane field configuration, we must
first have  a $d$-dimensional Poincar\'e symmetry for the space--time metric
$G_{MN}$ and the antisymmetric potential $B_{M_1 \cdots M_d}$. We
do not necessarily require a transverse $SO(D - d)$ symmetry at the beginning
for the metric and the antisymmetric potential, but at the end they do have
this symmetry. For simplicity,
we impose it from the beginning. As in \duflpb,   we make the most
general ansatz  satisfying the above requirements, for the metric
\eqn\genmetric{d S^2 = e^{2 A (r)} \eta_{\mu\nu} d x^\mu d x^\nu + e^{2 B (r)}
\delta_{mn} d y^m d y^n,}
and for the potential
\eqn\antisym{A_{01\cdots d-1} = - e^{C(r)},}
where $\mu = 0, 1, \cdots, d - 1; m = d, d + 1, \cdots, D -1$;
$r = \sqrt {\delta_{mn} y^m y^n}$ and $A(r), B(r)$ are as yet undetermined
$SO(D - d)$ invariant functions. All the other components of $G_{MN}$ and
$B_{{M_1} \dots {M_d}}$ are set to vanish. We choose the static gauge
\eqn\staticg{X^\mu = \xi^\mu.}
Since we want to find a {\it super} static ($d - 1$)-brane configuration,
we must have the supersymmetry transformations satisfied for $\theta = 0$.
This is effectively achieved by requiring the existence of some non-vanishing
Killing spinor $\epsilon$ such that
\eqn\susy{\delta \theta = (1 - \Gamma)\epsilon = 0,}
where we have used the local fermionic gauge symmetry \skappa\ to gauge away
{\it half} of the components of $\theta$. By using the static gauge
\staticg, \genmetric\ and \wmetric, we have the $\Gamma$ in \skappa,
\eqn\staticgamma{\Gamma = (- 1)^{(d - 2)(d + 1)/4} \Gamma_{01\cdots d - 1},}
where $\Gamma$ is  actually a $d$-dimensional chiral operator as $d$ is even.
Equation
\susy\ tells that {\it half} of the supersymmetries must be broken for this
static
super $(d - 1)$-brane field configuration, which is entirely consistent with
what we learned on finding  super extended solutions from supergravity
theories
\refs{\dabghr, \duflfb, \dufs,\calhswb, \duflthb, \duflbscan, \duflsds}.
Therefore,
 we expect that the field
configuration has a $d$-dimensional super Poincar\'e symmetry. We now consider
the world-volume equations of motion for vanishing $\theta$. They are actually
the same as those derived from the bosonic action \pbact, which are
\eqn\eqmotion{\eqalign{\partial_i \big(\sqrt {- \gamma}\gamma^{ij} \partial_j
X^N G_{MN}\big) -&\half \sqrt{- \gamma}\gamma^{ij}\partial_i X^N\partial_j X^P
\partial_M G_{NP} \qquad \qquad \qquad {}\cr
 - &{1\over d!} \epsilon^{i_1\cdots i_d}
\partial_{i_1} X^{N_1} \cdots \partial_{i_d} X^{N_d} F_{M{N_1}\cdots {N_d}}
= 0,\cr}}
and
\eqn\inducedmetric{\gamma_{ij} = \partial_i X^M \partial_j X^N G_{MN}.}
Using the static gauge \staticg\ and the ansatz for \genmetric\ and \antisym,
the above equations reduce to
\eqn\reducedeq{\partial_m \big(e^{d A} - e^C \big) = 0}
and
\eqn\redwmetric{\gamma_{\mu\nu} = e^{2 A} \eta_{\mu\nu}.}
Equation \reducedeq\ is the so-called ``no-force" condition, and the
term within the bracket is actually the potential (see \refs{\duflfb, \dufs}).
Hence eq. \reducedeq\ implies that the potential is constant, and
the $(d - 1)$-brane itself therefore does not feel any back-reaction. We can
always
choose the zero point of the potential so that the constant vanishes. Thus, we
have from \reducedeq,
\eqn\acrelation{C = d A.}
In the next section, we will determine the unknown parameter $\alpha (d)$,
and the functions $A$ and $B$.

\newsec{\bf Determination of the couplings}

In \duflpb, we found the ($d - 1$)-brane field configuration by imposing
$P_d \times SO(D - d)$ symmetry on the background fields simply from the
combined bosonic action \genactone\ + \pbact. The surprising thing is that
the unknown parameter $\alpha (d)$ is determined in the process of finding
solutions for $D > d + 2$, where $D$ the space--time dimension. We are not
going to repeat the same derivation here.
 Instead, we just quote the solitonic ($d - 1$)-brane solution for
an {\it arbitrary} $\alpha (d)$ from \duflpb, and demand  that this solitonic
configuration for $r\not = 0$ satisfy the world-volume equations of motion,
i.e.
the ``no-force" condition discussed in the previous section, then find the same
$\alpha (d)$. This is analogous to dealing with  Dirac's monopole,
which can be described either by a singular Dirac-string in a space with
trivial
topology or by Wu--Yang construction in a space with non-trivial topology. For
$r\not = 0$, both descriptions give the same field configuration. Things here
are a little  different. We do not need a source to find a magnetic or
topological $(d - 1)$-brane solution from the equations of
motion of the dual action of \genactone. This dual action is
\eqn\dualgenact{{\tilde I}_D = {1\over 2 \kappa^2}\int d^D x \sqrt{- g}
\bigg[ R - \half (\partial \phi )^2 - {1\over 2 (\tilde d + 1)!}
e^{- \alpha (\tilde d ) \phi} {\tilde F}_{\tilde d + 1}^2 \bigg] ,}
where $\alpha (\tilde d ) = - \alpha (d)$ and
\eqn\dualrelation{{\tilde F}_{\tilde d + 1} = e^{-\alpha (d)\phi}
{\,^*}F_{d + 1},}
where * is the Hodge dual operator.  We can find solitonic $(d - 1)$-brane
solutions from \dualgenact\ for an {\it arbitrary} $\alpha (d)$. However, the
Dirac-string-like,  so-called elementary solutions are obtained in
\duflpb\ only for fixed
$\alpha (d)$. The reason is that the sources are extended objects, described by
actions \pbact, some of which, i.e. those in {\bf O} and {\bf H} sequences, are
actually supersymmetric. This is one example to show that extended objects
differ from point particles. We expect that if we require that
the solitonic configurations for an {\it arbitrary} $\alpha (d)$ satisfy the
world-volume equations of motion, i.e. the ``no-force" condition discussed in
the previous section,  we can also fix the $\alpha (d)$. This is indeed the
case. We are about to demonstrate it explicitly.  As before, the most general
ansatz for the metric with $P_d \times SO(D - d)$ symmetry is
\eqn\canometric{ds^2 = e^{2a(r)}\eta_{\mu\nu} dx^\mu dx^\nu + e^{2 b (r)}
\delta_{mn} dy^m dy^n ,}
where notations have the same meanings as those in \genmetric, except that
$ds^2$ is a canonical metric. The relations between $ds^2, a$ and $b$ and
$dS^2, A$ and $B$ are given, by using \metrelation, as
\eqn\SABsab{\eqalign{dS^2 &= e^{{\alpha (d)\over d}\phi} ds^2 ,\cr
                        A &= a + {\alpha (d)\over 2d}, \cr
                        B &= b + {\alpha (d)\over 2d}. \cr}}
The solitonic ($d - 1$)-brane solution for an {\it arbitrary} $\alpha (d)$ from
\dualgenact\ is
\eqn\solitconfi{\eqalign{{\tilde F}_{\tilde d + 1} &= Q \epsilon_{\tilde d +
1},
\cr
ds^2  &= \bigg[1 + {|Q|\over \tilde d r^{\tilde d}}  \bigg]^{- \tilde d
\over D - 2} \eta_{\mu\nu} dx^\mu dx^\nu \cr
&\quad +\bigg[1 + {|Q|\over \tilde d r^{\tilde d}} \bigg]^{d\over D - 2}
(dr^2 + r^2 d\Omega_{\tilde d + 1}^2 ),\cr
e^{-{2\over \alpha (d)}\phi} &= 1 + {|Q|\over \tilde d r^{\tilde d}},
\cr}}
where we have set $\phi_0 = 0$ and $\epsilon_n$ is the volume
element of the unit $n$-sphere. By using \solitconfi\ and the dual relation
\dualrelation, we can read the ($d + 1$)-form $F_{d + 1}$ as
\eqn\newf{F_{r01\cdots d-1} = - {2\over \alpha} e^{(\alpha - {2\over \alpha}
+ {2d\tilde d\over \alpha (D - 2)})\phi} \partial_r \phi.}
By using the ``no-force" condition \acrelation, the relation \SABsab\ and
 the  metric $g_{MN}$ in \solitconfi, we can read the field strength \poten\
of potential $B_d$ as,
\eqn\oldf{\eqalign{F_{ro1\cdots d -1} &= - \partial_r e^C \cr
&= - \bigg({\alpha \over 2} + {d{\tilde d}\over \alpha (D - 2)}\bigg) e^{(
{\alpha \over 2} + {d\tilde d\over \alpha (D - 2)})\phi}\partial_r \phi.\cr}}
Identifying eq. \newf\ with \oldf\ for $r\not = 0$,  we have
\eqn\relalpha{{2\over \alpha (d)} = {\alpha (d)\over 2} + {d\tilde d \over
\alpha (d) (D - 2)}.}
Solving this simple equation, we have
\eqn\explialpha{\alpha^2 (d) = 4 - {2d\tilde d \over D - 2},}
which was already obtained  in \duflpb\ by finding $(d - 1)$-brane
solutions from \genactone\ + \pbact.  We expect that these
$\alpha (d)$ for $d \geq 2$ give the correct couplings in the corresponding
$N = 1$ supergravity theories, whose $(d, D)$ fall in either {\bf O} or {\bf H}
sequences. By inspection, this is indeed the case. These couplings  actually
cover all the dilaton--antisymmetric tensor ones except for those
containing 2-forms in $N = 1$ supergravity theories.  We determine the
$\alpha (d)$ simply by demanding the solitonic solution with an arbitrary
$\alpha (d)$ satisfying the world-volume equations of motion, i.e.
the ``no-force" condition. Therefore, it is determined by space--time
supersymmetry, since the static configuration in the previous section is
supersymmetric.  One may ask why  those $\alpha (1)$ for $d = 1$ in the same
sequences do not describe the right couplings, involving 2-forms in the
corresponding supergravity theories.
The simple answer is that we do not have equal on-shell matching of fermionic
and bosonic degrees of freedom for the
superparticle, which implies no world-line supersymmetry,  although we can have
both space--time supersymmetry and local $\kappa$-gauge symmetry. In order to
obtain
 the correct coupling, the crucial thing is that
we must have equal on-shell fermionic and bosonic degrees of freedom for our
field
configuration, which can be achieved by having the $\kappa$-symmetry for
a $p \geq 1$ extended object with space--time supersymmetries. However, this is
not
the whole story, we will come back to this point later on. Does our $\alpha
(d)$
apply to $N \geq 2$ supergravity theories? If so, what does this imply?
 This would be the topic of the next section.

\newsec{\bf On $N \geq 2$ super extended objects and supergravity theories}

By inspection, our $\alpha (d)$ for $d > 1$ also give the correct couplings in
Type II $D = 10$ supergravity theories. This is also true whenever we have a
coupling containing  an antisymmetric tensor  as a singlet of the internal
symmetry of any  given $D \leq 9, N \geq 2$ supergravity  multiplet.
 Careful inspection shows
that this actually happens only for  $N = 2$ supergravity theories.
This is entirely
consistent with the spin-content analysis, given in \calhswb, that we cannot
go beyond
spin $1$ on the world-volume; therefore we cannot go beyond Type II super
extended objects.
These results are quite unexpected. From our previous discussions, the
immediate conclusion we can draw is that the corresponding $N = 2$ super
$p$-brane actions, if they exist at all,  must reduce to the bosonic
$p$-brane actions \pbact\ after we set  the fermionic fields and
possible world-volume spin $1$  fields to vanish. At the same time, we break
only
{\it half} of the space--time supersymmetries.  We may also jump to the
conclusion
that there might exist supercovariant actions with $N = 2$ space--time
supersymmetries for all the aforementioned cases. This is possibly true
for only Type II $D = 10$  $p$-branes for $p \geq 0$.  We cannot have a
supercovariant action with a world-volume vector field for each of
$N = 2 , D \leq 9$ cases, since now there
is no equal on-shell matching of bosonic and fermionic degrees of freedom, as
discussed in \duflbscan.  Our results  not
 only support the spin-content analysis but also  provide a first step toward
constructing the complete Type II  super $p$-brane actions for $p > 1$.
But we still have quite a few unanswered questions on those lower-dimensional
supersymmetric soliton solutions obtained from dimensional and
double-dimensional reductions of higher-dimensional solutions. It is actually
the main topic of this section to understand the nature of all these solutions.
The first question is to know why our approach applies only to those
aforementioned
cases, and  not to all the soliton solutions classified in
\refs{\duflpb, \duflbscan}. For concreteness, we take $N = 2, D = 8$
supergravity \salstwo\ as an example. Its field content is
\eqn\fieldcontent{g_{MN},\quad 2 \psi_M,\quad 6 \chi, \quad A_{MNP},\quad
3 A_{MN}, \quad 6 A_M, \quad 7 \phi,}
where spinors are {\it pseudo-Majorana} and the bosons are real. The
automorphism group of the superalgebra is $SU(2)$. With respect to this
$SU(2)$,
$\psi_\mu^a$ ($a = 1, 2$) is a 2-spinor, $\chi_i^a$ is a vector-spinor,
$A_{\mu\nu}^i$ is a triplet, and the vectors split into two triplets. Besides,
there is another internal-symmetry
 ${\rm global} SL(3, R) \times  {\rm local} SO(3)$, and
the down and up indices $i$ are also in the representations 3 and $\bar 3$ of
global $SL(3, R)$, respectively. The seven scalars parametrize the coset
$SL(3, R)/SO(3)$ and transform as 5 + 1 + 1 under the $SU(2)$.
All the supersymmetric solutions of Type IIA $D = 10$ supergravity,  which are
independent of more than one spatial dimension, must also be those of the
supergravity defined above, since the latter can be obtained from the former
 by a dimensional reduction on
 $S^1 \times S^1$. As discussed in \duflpb, any  solution with the
bosonic symmetry $P_d \times SO(D - n)\times (S^1)^{n - d}$
$(D -3 \geq n \geq d)$
and breaking {\it half} of
the space--time supersymmetries guarantees solutions with bosonic symmetries
$P_k \times SO(D - l) \times (S^1)^{l - k}$ with $d \geq  k > 0$ and
$n \leq l \leq D - 3$ and also breaking  {\it half} of the space--time
supersymmetries.
 Therefore, we
should have six identical supersymmetric-particle solutions corresponding to
six 1-form potentials, three identical string solutions to three 2-form
potentials, and one
membrane to one 3-form potential in the supermultiplet. We have the same number
of solutions for the dual object of each of the above cases. All these
solutions
break half of the space--time supersymmetries and have the corresponding
maximal bosonic
symmetries: $P_1 \times SO(7)$ for superparticle, $P_2 \times SO(6)$ for
superstring, $\cdots$, and
$P_5 \times SO(3)$ for super 4-brane. However, these solutions have no
fundamental world-volume actions but gauge-fixed ones by dimensional and
double-dimensional reductions of Type II $D = 10$ super $p$-branes.
Each of these solutions guarantees
those  with smaller bosonic symmetries, as described  above. Out of these
solutions, our approach applies only to the membrane solutions since only the
3-form is a singlet of the internal symmetry. Our $\alpha (d)$ for
$d = 3, D = 8$ indeed gives the correct coupling. Let us explain it in detail
and take the 2-forms $A_{MN}^i$ as examples. We have three string solutions
with the maximal bosonic symmetry $P_2 \times SO(6)$, one of which comes from a
Type IIA $D = 10$ superstring by dimensional reduction on $S^1 \times S^1$,
and the other two come from a Type IIA supermembrane by different routes.
One route is first to dimensionally  reduce the $D = 10$ membrane to $D = 9$,
then to $D = 8$ by double-dimensional reduction. The other switches just the
order of dimensional and double-dimensional reductions. From the $D = 10$ point
of view, a Type II string in $D = 8$ is either a string or a membrane in
$D = 10$.
Therefore, we should not expect to find simultaneously all  three strings
preserving the maximal bosonic symmetry and breaking just half of space--time
supersymmetries: were it the case, we would find simultaneously a
Type IIA superstring and two Type IIA supermembranes in $D = 10$, but just
breaking half of the space--time supersymmetries. It is obviously impossible.
So we can find at most two strings at one time from $D = 8, N =2$ supergravity.
Inspecting the equations of motion and the SUSY transformation rules, we indeed
find  the case. All these imply that if we want to find string solutions
breaking just {\it half} of
the space--time supersymmetries, we must pick a specific direction of the
internal symmetry, i.e. break the internal symmetry down to
 ${\rm global} U(1) \times {\rm global} SL(2, R) \times{\rm local} SO(2)$.
This in turn implies that the $D =10$ dilaton field for the string solution
is not a singlet of the internal symmetries of $D = 8, N =2$ supergravity.
Therefore, our approach does not apply to this kind of solutions since
the space--time metric $g_{MN}$ in a given dimension $D$ is always a singlet of
the internal symmetry of a given supergravity theory{\footnote \at {More
precisely, our discussion in section 2 does not apply to this kind of solution,
 but
our approach may still do.}. However, the membrane
solution has nothing to do with the internal symmetry, the $D =10$ dilaton
can now be identified with the $D = 8$ one, and our approach
works. Indeed, this is  the general feature of our approach in all supergravity
\vfill\eject
\noindent
theories, and  explains why it always works, except for
2-form field strength in each of Type I cases.

 We conclude that {\it our approach always gives
the correct coupling if the corresponding antisymmetric gauge potential is
a singlet of the internal symmetry of a certain  supergravity theory and there
exists  either a fundamental or a gauge-fixed
super $p$-brane action ($p \geq 1$) consistently coupled with this potential}.

We now come to
talk about superparticles.
As discussed in the previous section, our approach does not, in general, apply
to these. However, one may notice already that our $\alpha (d)$ does
give the correct coupling for $d = 1$ in Type IIA $D = 10$ supergravity. The
determination of this $\alpha (1)$ is actually not through the superparticle,
but its dual object, a super 6-brane for which we have equal on-shell matching
 of
bosonic and fermionic degrees of freedom. Hence besides those conditions
mentioned above for our approach, if there exists a super $p$-brane with
$p \geq 1$ and the field strength of its antisymmetric background field is dual
to a 2-form, then the $\alpha$ associated with the 2-form is correctly
determined and must be with opposite sign to that of the $p$-brane. At the same
time, we know that there exists also a superparticle solution from the
corresponding supergravity. By inspection, our $\alpha (d)$ for $d = 1$ also
give  correct couplings for $N =2, D = 9$ and $N = 2, D = 5$ supergravity
theories. This suggests that we might have $N = 2$ superparticles in these
dimensions. By counting, we do have equal on-shell matching of fermionic and
bosonic
degrees of freedom for these superparticles. Further investigations show that
equal on-shell matching of bosonic and fermionic degrees of freedom happens
also
for $N = 2$ superparticles in $D = 3$ and $D =2$. These superparticles, if they
exist at all, must share all the properties of  Type I super $p$-branes.
Therefore, we expect that they can be obtained from $N = 2$ superstrings by
double-dimensional reductions described in  \dufhis, and their world-volume
actions are just those given in \berst\ for $p = 0$.

We have one last thing to say in this section, i.e. that our above discussion
does not apply to self-dual cases, since we did not incorporate anything
on self-dual properties.  Thus we should not expect that our
$\alpha (d)$ gives the right couplings for those cases.
As we know, there exist only two cases, and
only for Type II supergravity theories: one is $D = 10$ Type IIB supergravity
with a self-dual 5-form, the other is $D = 6$ pure Type IIB supergravity with
a self-dual 3-form. The former one should correspond to a self-dual Type IIB
super 3-brane \duflthb, and the latter to a self-dual Type IIB superstring
\duflsds.  The bosonic sector of the self-dual Type IIB superstring should also
be given by \pbact, but with an additional self-dual relation for the 3-form.
Our previous discussions tell that the bosonic part of Type II super $p$-brane,
 after setting zero fermionic coordinates and
world-volume spin $1$ field, is also given by \pbact.
 Therefore, the same situation as
for the self-dual string may apply to the self-dual Type IIB super 3-brane.
It is very easy to get the answers. In both cases, we should no longer use the
 action
\genactone\ because of the self-dual relations for 3-form and 5-form.
However, the equations of motion derived for graviton and dilaton are still
valid. For the self-dual string, these equations should be identical, whether
 we use $F_3$ or $^* F_3$. More precisely, the $\alpha$ should be the same for
either $F_3$ or $^* F_3$. This immediately leads to
$\alpha (3) = -\alpha (3)$, i.e. $\alpha (3) = 0$. The
same applies to the 5-form, so $\alpha (5) = 0$. Hence for self-dual cases,
$\alpha (d) = 0$ even without referring to solutions. It is worth pointing
out that $\alpha (5) = 0$  also fits into eq. \relalpha. We do not know
if there is a deep physical reason behind this.

\newsec{\bf Space--time supersymmetry}

Once the action \genactone\ is determined, we can use it to find
Reissner--Nordstr\"om-like solutions. Since we know that the action \genactone\
is part of the bosonic sector of some supergravity action for each $\alpha$,
the charge = mass
limit of the corresponding Reissner--Nordstr\"om-like solution must be the
supersymmetric configuration that solves both the equations of motion and the
supersymmetric transformation rules for vanishing gravitino $\psi_M$ and
dilatino $\lambda$, of the corresponding supergravity theory. We will use this
fact to determine the supersymmetric transformation rules for vanishing
$\psi_M$
and $\lambda$. After doing this, we have reached a  stage in the supergravity
action, which is good for studying almost all the vacuum-like solutions of the
supergravity theory that are related to the determined couplings. The
general Reissner--Nordstr\"om-like extended solutions
of \genactone, in diverse dimensions, have been given in \duflpb\ for the
$\alpha(d)$ given by \explialpha. Their charge = mass limits have also been
given there; they are quoted as:
\eqn\superconf{\eqalign{ a = {\tilde d\over 2(D - 2)}C,\quad &  \quad
b = -{d \over 2 (D - 2)}C, \cr
B_{01\cdots d-1} = - e^C ,\quad &\quad  \phi = {\alpha (d) \over 2}C,\cr}}
and
\eqn\explic{ e^{- C} = 1 + {|Q|\over \tilde d r^{\tilde d}} ,}
where $D$ is the space--time dimension, $a$ and $b$ are defined by
\canometric, and
we have set $C_0 = 0$ above. Since fermionic fields differ for different
supergravity multiplets, and  could be Dirac, Majorana, Weyl or Majorana--Weyl,
etc., the most general supersymmetric transformation rule for a given
spinor depends crucially on its kind. Hence we have to deal with each
supergravity multiplet separately. In what follows, we just give a familiar
example, i.e. the $D = 10, N = 1$ supergravity with 3-form field strength, to
demonstrate the procedure. For this supergravity, both gravitino and dilatino
are Majorana--Weyl, but with opposite chirality.  The most general
supersymmetric transformation rules for vanishing $\psi_M$ and $\lambda$ are
\eqn\susytransf{\eqalign{\delta \lambda &= \Gamma^M \partial_M \phi \epsilon +
g(\phi) \Gamma^{MNP}F_{MNP}\epsilon = 0 ,\cr
\delta \psi_M &= D_M \epsilon + h(\phi) (\Gamma_M\,^{NPQ} - t \delta_M\,^N
\Gamma^{PQ})F_{NPQ}\epsilon = 0,\cr}}
where we have dropped the overall factors and set the $D = 10$ gravitational
constant $\kappa = 1$; $g(\phi)$ and $h(\phi)$ are two as yet undetermined
functions of the dilaton field, $t$ is an undetermined numerical constant, and
 $\epsilon$ is the $D = 10$ Majorana--Weyl spinoral parameter, which satisfies
\eqn\Weylcond{\Gamma_{11}\epsilon = \Gamma_{11},}
where $\Gamma_{11}$ is defined as
\eqn\defoneone{\Gamma_{11} \equiv \Gamma_0 \Gamma_1 \cdots \Gamma_9,}
with flat indices $0, 1, \cdots, 9$.
We first use the scaling arguments described in section 2 to determine the
functions
$g(\phi)$ and $h(\phi)$. We expect that each of the equations in \susytransf\
should scale homogeneously under the following scalings
\eqn\bgscalings{\eqalign{g_{MN} &\rightarrow \lambda^{1/2} g_{MN},\cr
             e^\phi &\rightarrow \lambda^{3/ \alpha (2)} e^\phi, \cr
             B_{MN} &\rightarrow \lambda^2 B_{MN} , \cr}}
which are obtained by setting $D = 10, d = 2$ from the general scalings
\genscalings\ in
section 2. Since the $\alpha (d)$ is determined up to its square, however, we
can always choose the sign for the $\alpha (d)$ if the considered supergravity
multiplet involves only one antisymmetric tensor $F_{d + 1}$. The reason is
that we can always get the chosen sign for the $\alpha (d)$ by sending
$\phi \rightarrow - \phi$ in the action \genactone. But for supergravity
involving
more than one antisymmetric tensor, we must take care of the relative signs
of those $\alpha$'s. This will be the topic of the next section. In what
follows, we simply choose the positive sign, i.e. from \explialpha,
$\alpha (2) = 1$. By noting that
$\Gamma_M \rightarrow \lambda^{1\over 4} \Gamma_M$
in \susytransf, it is easy to know that $g(\phi)$ and $h(\phi)$ must scale as
\eqn\ghscaling{\eqalign{g(\phi) &\rightarrow \lambda^{-3/ 2} g(\phi), \cr
                        h(\phi) &\rightarrow \lambda^{-3/ 2} h(\phi). \cr}}
Comparing  with the scaling of $e^\phi$ in \bgscalings, we have
\eqn\ghform{\eqalign{g(\phi) &= g_0 e^{- \phi/2} ,\cr
                     h(\phi) &= h_0 e^{- \phi /2},\cr}}
where $g_0$ and $h_0$ are as yet two undetermined constants. We proceed by
substituting \superconf\ for $D = 10, d= 2$ to \susytransf. The results are
\eqn\simpsusy{\eqalign{\delta \lambda &= \half \Gamma^m \partial_m C (1 +
2\cdot 3! g_0 \Gamma_{01})\epsilon = 0,\cr
\delta \psi_\mu &= \partial_\mu \epsilon - {3\over 16} \Gamma^n\,_\mu
\partial_n C (1 - {32\over 3} t\cdot h_0 \Gamma_{01})\epsilon = 0, \cr
\delta \psi_m &= \partial_m \epsilon - 2 t\cdot h_0 \Gamma_{01} \partial_m C
\cr
&\quad + {1\over 16} \partial_n C \Gamma^n\,_m (1 - 16\cdot 3! h_0 \Gamma_{01})
\epsilon = 0,\cr}}
where all indices are world ones except for the flat $0,1$ in the
$\Gamma_{01}$.
By \susy\ and \staticgamma, as discussed in section 3 from the
world-volume point of view, the supersymmetric configuration \superconf\ must
break half of the space--time supersymmetries. For the present case, it is
equivalent to saying
\eqn\halfsusy{(1 - \Gamma_{01})\epsilon = 0,}
where indices 0 and 1 are flat, too. Combining \simpsusy\
with \halfsusy, we  have
\eqn\ght{g_0 =- {1\over 12},\quad h_0 = {1\over 96},\quad t = 9.}
Inserting the above $g_0, h_0$ and $t$ back into \susytransf, the resulting
transformation rules for vanishing $\psi_M$ and $\lambda$ agree perfectly
with their correspondents derived either by  direct construction or by
dimensional reduction and truncation from $D = 11$ supergravity.

\newsec{\bf  Cases involving  many antisymmetric tensors}

For a supergravity multiplet involving many antisymmetric tensors,
  we first determine  the action and
supersymmetric transformation rules for each of the antisymmetric tensors as
described in the previous
sections. Then, the action and supersymmetric transformation rules
involving many antisymmetric tensors are easily obtained:  the
single antisymmetric tensor contribution in \genactone\ and  the corresponding
supersymmetric transformation rules are replaced by the  contributions
from all antisymmetric tensors. In this way, we have constructed a
supergravity action, and its SUSY
transformation rules, for vanishing fermionic fields, up to such a stage that
it
is useful for almost all the vacuum-like solutions of the full supergravity
theory. However, we still have one unsolved problem, as mentioned earlier, i.e.
we do not know how to determine the relative signs of those $\alpha$'s in
both the action and SUSY transformation rules. Actually, there exist only two
such supermultiplets to which our approach applies, and both of them live in
$D = 10$: one is Type IIA supergravity, the other Type IIB. We do not have any
problem with the Type IIB one, since we have only two $\alpha$'s and one is
zero.
So we need to deal with the Type IIA one only. Up to relative signs of the
$\alpha$'s, the determined Type IIA action
and supersymmetric transformation rules for vanishing gravitino $\psi_M$ and
dilatino $\lambda$ are
\eqn\typeiiact{\eqalign{I_{10} ({\rm Type \quad IIA}) &= {1\over 2 \kappa^2}
\int d^{10} x \sqrt{-g} \bigg[R - \half (\partial \phi)^2 - {1\over 2\cdot 2!}
e^{{- \alpha (1)}\phi} F_2^2 \cr
&\quad -{1\over 2\cdot 3!} e^{{- \alpha (2)}\phi} F_3^2 - {1\over 4!}
e^{{-\alpha(3)}\phi}F_4^2 \bigg],\cr}}
and
\eqn\iilambda{\eqalign{\delta \lambda &= \Gamma^M \phi \Gamma^{11} \epsilon
-{\alpha (1)\over 4} e^{-\alpha (1)/ 2} \Gamma^{NP} F_{NP}\epsilon \cr
&\quad + {i \alpha(2) \over 12} e^{-\alpha (2)\over 2} \Gamma^{NPQ} F_{NPQ}
\epsilon \cr
&\quad + {i \alpha(3) \over 48} e^{-\alpha(3)/ 2} \Gamma^{NPQR} F_{NPQR}
\epsilon ,\cr}}
\eqn\iipsi{\eqalign{\delta \psi_M &= D_M \epsilon + {1\over 64}
e^{-\alpha (1)/ 2} (\Gamma_M\,^{NP} - 14 \delta_M\,^N \Gamma^P )F_{NP}
\Gamma^{11} \epsilon \cr
&\quad + {1\over 96} e^{-\alpha (2)/ 2} (\Gamma_M\,^{NPQ} - 9 \delta_M\,^N
\Gamma^{PQ} )F_{NPQ} \Gamma^{11} \epsilon \cr
&\quad + {i\over 256} e^{-\alpha (3)/ 2} (\Gamma_M\,^{NPQR} - {20\over 3}
\delta_M\,^N \Gamma^{PQR} )F_{NPQR} \epsilon,\cr}}
where, from \explialpha,
\eqn\thosealpha{\alpha^2 (1) = {9\over 4},\quad \alpha^2 (2) = 1, \quad
\alpha^2 (3) = {1\over 4},}
where $\Gamma^M$ are the $D = 10$ Dirac matrices, where the covariant
derivative is given by
\eqn\coderi{D_M = \partial_M + {1\over 4}\omega_{MAB} \Gamma^{AB}, }
with $\omega_{MAB}$ the Lorentz spin connection, where
\eqn\gamatri{\Gamma^{N_1 N_2 \cdots N_n} = \Gamma^{[N_1} \Gamma^{N_2} \cdots
\Gamma^{N_n ]}}
and where
\eqn\gammaeleven{\Gamma^{11} = i \Gamma^0 \Gamma^1 \cdots \Gamma^9,}
with flat indices $0, 1, \cdots, 9$. It seems that our approach does not
provide a way to determine the relative signs of the $\alpha$'s in \thosealpha.
 Here
we appeal to string quantum-loop arguments. We
know that this supergravity is the field theory limit of Type IIA superstring.
If we write the action \typeiiact\ in string $\sigma$-model variables, each
term
in the action should correspond to a positive or at least a tree-level string
quantum loop, because we have a good  superstring quantum theory. Since the
overall sign for those $\alpha$'s is not important classically, for simplicity,
we choose $\alpha (2) = 1$ from $\alpha^2 (2) = 1$. Writing the action
\typeiiact\ in string $\sigma$-model variables, i.e. from \metrelation\ and
\explialpha\ for $d = 2, D = 10$,
\eqn\smetrelat{G_{MN} = e^{\phi/2} g_{MN},}
we have
\eqn\typeiisigma{\eqalign{I_{10} ({\rm Type \quad IIA}) &= {1\over 2 \kappa^2}
\int d^{10} x \sqrt{-G} e^{-2\phi} \bigg[R + 4 (\partial \phi)^2 -
{1\over 2\cdot 2!} e^{{1 - 2 \alpha (1)\over 2}\phi} F_2^2 \cr
&\quad -{1\over 2\cdot 3!}  F_3^2 - {1\over 4!} e^{{3 -2\alpha(3)\over 2}\phi}
F_4^2 \bigg],\cr}}
where we can read the string loop-counting parameter as $e^\phi$. Hence
 both ${1 - 2 \alpha (1)/ 4}$ and ${3 - 2\alpha (3) / 4}$  must be
integers $\geq 0$. From \thosealpha, this is only possible for
$\alpha (1) = - 3/2$ and  $\alpha (3) = -1/2$. Inserting those $\alpha$'s
back into \typeiiact, \iilambda\ and \iipsi, we get almost the same as those
obtained by dimensional reduction from $D = 11$ supergravity with vanishing
fermionic fields. Our approach does not fix the $WZ$ term and those modified
terms in $F_2, F_3$ and $F_4$ in the Type IIA bosonic
action, since they vanish for our static configurations.

\newsec{\bf A prediction of $D = 9, N = 2$ supergravity}

The $D = 9, N = 2$ supergravity has field content
\eqn\sugracont{g_{MN},\quad 2\psi_M , \quad A_{MNP}, \quad 2 B_{MN}, \quad
3 A_M, \quad 4 \chi, \quad 3\phi.}
The three scalars parametrize the coset $GL(2, R)/SO(2)$. The spinors are
pseudo-Majorana. This theory has not been written down so far because it is
tedious, even though trivial,
 to obtain by dimensional reduction from $D = 11$
supergravity. Since the 3-form potential $A_{MNP}$ is a singlet of
internal symmetry $GL(2, R)\times SO(2)$, and the corresponding supermembrane
action must be the gauge-fixed one of the $D = 11$ supermembrane, we should
be able to determine, by our approach, the corresponding coupling and
supersymmetric
transformation rules for vanishing fermionic fields.  From
our formulae \explialpha\ and \genactone, the coupling should be
\eqn\predcoupling{I_9 (N = 2) = {1\over 2 \kappa^2} \int d^9 x \sqrt {- g}
\bigg[R - \half (\partial \phi)^2 - {1 \over 2\cdot 3!}
e^{-{ 2\over \sqrt 7}\phi}F_4^2 \bigg],}
where $F_4 = d A_3$.
The determined SUSY transformation rules for vanishing fermionic fields and
bosonic fields other than those in \predcoupling\ are
\eqn\presusy{\eqalign{\delta \chi_i &= \Gamma_i \Gamma^M \partial_M \phi
\epsilon + {1\over 4! \cdot \sqrt 7} e^{- {\phi \over \sqrt 7}} \Gamma_i
\Gamma^{NPQR} F_{NPQR} \epsilon ,\cr
\delta \psi_M &= D_M \epsilon - {1\over 224} e^{-{\phi \over \sqrt 7}} \bigg(
\Gamma_M\,^{NPQR} - {16\over 3} \delta_M\,^N \Gamma^{PQR} \bigg) F_{NPQR}
\epsilon , \cr}}
where $M = 0, 1, \cdots, 8$ and $i = 1, 2$ are $SO(2)$ vector indices,
where $D_M$ is given by
\eqn\dcovariant{D_M = \partial_M + {1\over 4} \omega_{MAB} \Gamma^{AB},}
with $\omega$ the Lorentz spin connection, and where $\Gamma^i$ and $\Gamma^A$
are given by
\eqn\diracmatrix{\Gamma^i = 1 \times \sigma^i, \qquad \Gamma^A = \gamma^A
\times \sigma^3,}
with $\gamma^A$ the $D = 9$ Dirac matrices and $\sigma^1, \sigma^2$ and
$\sigma^3$ the Pauli matrices. These transformation rules are derived along the
same line as described in section 6. In what follows, we will perform
a dimensional reduction on the action of $D = 11$ supergravity on two tori
$S^1 \times S^1$ to confirm the coupling in \predcoupling, but we will not
touch
the SUSY transformation rules in eq. \presusy\ and leave them  as a prediction.

In order to have
\eqn\eletonine{\sqrt {- g_{11}} R_{11} = \sqrt {- g_9} R_9  + \cdots ,}
the relation between the $D = 11$ metric $g_{11}$ and the $D = 9$ canonical
metric $g_9$ has to be
\eqn\emetonme{g_{11} = \pmatrix{\varphi^{-{2\over 7}} g_9 &&\cr
                                  &\varphi & \cr
                                  &  & \varphi \cr},}
where $R_{11}$ and $R_9$ are $D =11$  and $D = 9$ Ricci scalars,
respectively, and the $\varphi$ is essentially the dilaton in \predcoupling.
Using \emetonme, the precise relation of \eletonine\ is
\eqn\precise{\sqrt {- g_{11}} R_{11} = \sqrt {- g_{11}} \bigg[R_9 -
{9\over 2\cdot 7} \bigg({\partial \varphi \over \varphi}\bigg)^2 \bigg].}
In order to compare \precise\  with \predcoupling, we have to take the kinetic
term
of the dilaton field with standard normalization factor $1/2$, which is
achieved by setting
\eqn\standard{\varphi = e^{- {\sqrt 7 \over 3}\phi}.}
By using \emetonme\ and \standard, it is easy to show that
\eqn\fcoupling{\sqrt {- g_{11}} F_4^2 (11) \rightarrow e^{-{2\over \sqrt
7}\phi}
\sqrt {-g_9} F_4^2 (9) ,}
where $F_4^2 (11)$ is the square of the rank-4 antisymmetric tensor of $D = 11$
supergravity, and $F_4^2 (9)$ is the correspondence of $D = 9, N = 2$
supergravity. It is easy to see that our prediction of the coupling in
\predcoupling\ is confirmed.

\newsec{\bf Conclusion}
 We have made it clear when and why the dilaton--antisymmetric tensor couplings
in the supergravity theories can be determined by our approach. The most
important feature of this approach is the equal on-shell matching of bosonic
and fermionic degrees of freedom, which implies supersymmetries both on
space--time and on world-volume through the so-called $\kappa$-symmetry.
Therefore, we should
not be so surprised by the determined couplings since they are determined in
supergravity theories by supersymmetry, too. Our results indicate also that
there might exist fundamental Type II $p$-branes. The known actions of the Type
I super $p$-branes may provide a starting point toward constructing those of
the recently classified Type II $p$-branes \refs{ \duflpb, \duflbscan}. This
paper provides just a primary step to spell out supergravity theories in the
sense that we have considered only the simplest field configurations. We
expect more, if sophisticated supersymmetric field configurations are
considered.
In turn, we may also benefit from this for understanding  the Type II
$p$-branes better.

\bigbreak\bigskip\bigskip\centerline{{\bf Acknowledgements}}\nobreak The author
 would like to express thanks to M. J. Duff, C. Kounnas and F. Quevedo   for
discussions, and to P. Howe for  the help on understanding the relation between
the equations of motion and the superspace constraints in supergravity
theories.

\vskip 2in
\listrefs
\bye